\documentclass{article}
\usepackage{spconf,amsmath,graphicx}
\usepackage{tabularx}
\usepackage{multirow}
\usepackage{color,soul}
\usepackage{hhline}

\newcommand\blfootnote[1]{%
  \begingroup
  \renewcommand\thefootnote{}\footnote{#1}%
  \addtocounter{footnote}{-1}%
  \endgroup
}

\title{MULTIMODAL EMOTION RECOGNITION WITH HIGH-LEVEL SPEECH AND TEXT FEATURES}

\name{Mariana Rodrigues Makiuchi, Kuniaki Uto, Koichi Shinoda}
\address{Tokyo Institute of Technology}
\begin{document}
%
\maketitle

\begin{abstract}


Automatic emotion recognition is one of the central concerns of the Human-Computer Interaction field as it can bridge the gap between humans and machines. Current works train deep learning models on low-level data representations to solve the emotion recognition task. Since emotion datasets often have a limited amount of data, these approaches may suffer from overfitting, and they may learn based on superficial cues. To address these issues, we propose a novel cross-representation speech model, inspired by disentanglement representation learning, to perform emotion recognition on wav2vec 2.0 speech features. We also train a CNN-based model to recognize emotions from text features extracted with Transformer-based models. We further combine the speech-based and text-based results with a score fusion approach. Our method is evaluated on the IEMOCAP dataset in a 4-class classification problem, and it surpasses current works on speech-only, text-only, and multimodal emotion recognition. \blfootnote{Copyright 2021 IEEE. Published in the 2021 IEEE Automatic Speech Recognition and Understanding Workshop (ASRU) (ASRU 2021), scheduled for 14-18 December 2021 in Cartagena, Colombia. Personal use of this material is permitted. However, permission to reprint/republish this material for advertising or promotional purposes or for creating new collective works for resale or redistribution to servers or lists, or to reuse any copyrighted component of this work in other works, must be obtained from the IEEE. Contact: Manager, Copyrights and Permissions / IEEE Service Center / 445 Hoes Lane / P.O. Box 1331 / Piscataway, NJ 08855-1331, USA. Telephone: + Intl. 908-562-3966.}
\end{abstract}
\section{Introduction}
\label{sec:introduction}


Correctly perceiving other people's emotion is one of the key components of good interpersonal communication. Emotions make conversation more natural. They can add or remove ambiguity, and they can change the meaning of what is being communicated altogether. Due to the importance of emotions in human-to-human conversation, automatic emotion recognition has been one of the main concerns of the Human-Computer Interaction (HCI) field for decades~\cite{levinson1986continuously}.

Many studies have proposed emotion recognition methods from para-linguistic features or from transcribed text data~\cite{wang2018sentiment, krishna2020multimodal, pan2012speech}. However, depending on paralanguage, the semantics of the linguistic communication may change and vice-versa. Thus, spoken text may have different interpretations depending on the speech intonation. Additionally, similar speech intonations may be used to convey different emotions, which can only be discerned by understanding the linguistic factor of communication. Therefore, relying solely on linguistic or para-linguistic information may not be enough to correctly recognize emotions in conversation.

One of the most fundamental challenges in emotion recognition is the definition of features that can capture emotion cues in the data. There is no agreement on the set of features that is the most powerful in distinguishing between emotions~\cite{kandali2008emotion, han2014speech, eyben2010opensmile}, and, in speech emotion recognition, this challenge is more aggravated, due to the acoustic variability introduced by speakers, speaking styles, and speaking rates.

Most current studies propose training deep learning models to extract those feature sets from the data~\cite{yeh2020speech, feng2020end}. Although these approaches have yielded satisfactory results, two problems remain. First, the training may easily lead to overfitting, since these models are usually trained from scratch using low-level data representations, and emotion recognition datasets are known to have a limited amount of data. Second, it is known that deep learning architectures may learn from superficial cues~\cite{goodfellow2014explaining}, which makes us question if current models can actually capture emotion information in the data.

We address these issues by challenging commonly used low-level data representations in speech-based and text-based emotion recognition studies, and by incorporating high-level features to our method. We define as low-level the features obtained from feature engineering, and we define as high-level the generic features extracted from deep learning approaches.

We propose a cross-representation model for speech emotion recognition, in which we aim to reconstruct low-level mel-spectrogram speech representations from high-level wav2vec 2.0 ones, thus leveraging both representations. We choose wav2vec because they contain rich prosodic information~\cite{pepino2021emotion}. Additionally, since we would like to capture a generic representation of emotion from speech, our model uses disentanglement representation learning techniques to eliminate speaker identity and phonetic variations in the data.
In our method, we also define a CNN-based model for text-based emotion recognition on features acquired from Transformed-based models. We believe these features are better than the commonly used word2vec and GloVe features due to the Transformer's ability of modelling long contextual information in a sentence, which is necessary for emotion recognition. Finally, we combine the results from the speech-based and the text-based models to obtain the multimodal emotion recognition results.


\section{Related Works}
\label{sec:related_works}

\subsection{Wav2vec 2.0}

Wav2vec 2.0~\cite{baevski2020wav2vec} is a framework to obtain speech representations via self-supervision. The wav2vec model is trained on large amounts of unlabelled speech data, and then it is fine-tuned on labelled data for Automatic Speech Recognition (ASR). Wav2vec is composed of a feature encoder and a context network. The encoder takes a raw waveform as input, and outputs a sequence of features with stride of $20$~ms and receptive field of $25$~ms. These features encode the speech's local information, and they have a size of $768$ and $1024$ for the ``base'' and ``large'' versions of wav2vec, respectively. The feature sequence is then inputted to the Transformer-based context network, which outputs a contextualized representation of speech. In the ``base'' and ``large'' wav2vec, there are $12$ and $24$ Transformer blocks, respectively. Although the wav2vec 2.0 learned representations were originally applied to ASR, other tasks, such as speech emotion recognition, can also benefit from these representations~\cite{pepino2021emotion, siriwardhana20, macary2021use}.



\subsection{Speech Emotion Recognition}
\label{sec:related_works-sub:ser}


During its early stage, most works on Speech Emotion Recognition (SER) proposed solutions based on Hidden Markov Models (HMM)~\cite{nwe2003speech}, Support Vector Machines (SVM)~\cite{pan2012speech, lin2005speech}, or Gaussian Mixture Models (GMM)~\cite{kandali2008emotion}. However, given the superior performance of deep learning on many speech-related tasks~\cite{chan2016listen,oord2016wavenet}, deep learning approaches for SER became predominant.



A problem characteristic to SER is the definition of appropriate features to represent emotion from speech~\cite{el2011survey}. Previous studies have attempted to extract emotion information from Mel-frequency cepstral coefficients (MFCC), pitch and energy~\cite{kandali2008emotion, han2014speech, eyben2010opensmile}. However, recent studies showed that employing a weighted sum with learnable weights to combine the local and contextualized outputs of a pre-trained wav2vec 2.0 model yields better speech emotion recognition results~\cite{pepino2021emotion}.






\subsection{Text Emotion Recognition}
\label{sec:related_works-sub:ter}

Current Text Emotion Recognition (TER) works use either features from an ASR model trained from scratch~\cite{feng2020end}, or Word2Vec or GloVe~\cite{krishna2020multimodal} features. Such works yield good results, but, given the outstanding performance of Transformer-based models in various NLP tasks~\cite{devlin2018bert, yang2019xlnet, clark2020electra}, it is natural to question commonly used representations for the TER task.

\subsection{Disentanglement Representation Learning}
\label{sec:related_works-sub:disent}




Disentanglement Representation Learning aims to separate the underlying factors of variation in the data~\cite{goodfellow2016deep}. The idea is that, by disentangling these factors, we can discard the factors that are uninformative to the task that we would like to solve, while keeping the relevant factors. Disentanglement has been applied to image~\cite{liu2018exploring}, video~\cite{denton2017unsupervised} and speech~\cite{qian2020unsupervised} applications. In speech-related works, it has been applied mainly to speech conversion and prosody transfer tasks~\cite{qian2019autovc,skerry2018towards}.

AutoVC~\cite{qian2019autovc} is an autoencoder that extracts a speaker-independent representation of speech content for speech conversion. A mel-spectrogram is inputted to the model's encoder, and the decoder reconstructs the spectrogram from the encoder's output and a speaker identity embedding. By controlling the encoder's bottleneck size, speaker identity information is eliminated at the encoder's bottleneck. SpeechFlow~\cite{qian2020unsupervised} builds upon AutoVC to disentangle speech into pitch, rhythm and content features.


Even though these works show impressive results in speech conversion, only few works attempt to disentangle speech for SER. \cite{williams2019disentangling} proposed an autoencoder to disentangle speech style and speaker identity from i-vectors and x-vectors, and they used the speech style embedding for SER. \cite{li2020speaker} used adversarial training to disentangle speech features into speaker identity and emotion features. These methods hold similarities with the SER method proposed in this paper, but our approach differs from previous works in that we explicitly eliminate speaker identity information from speech to obtain emotion features, and we perform experiments on the disentanglement property of these features.

\section{Methodology}
\label{sec:methodology}

\begin{figure*}[t]
  \centering
  \includegraphics[scale=0.29]{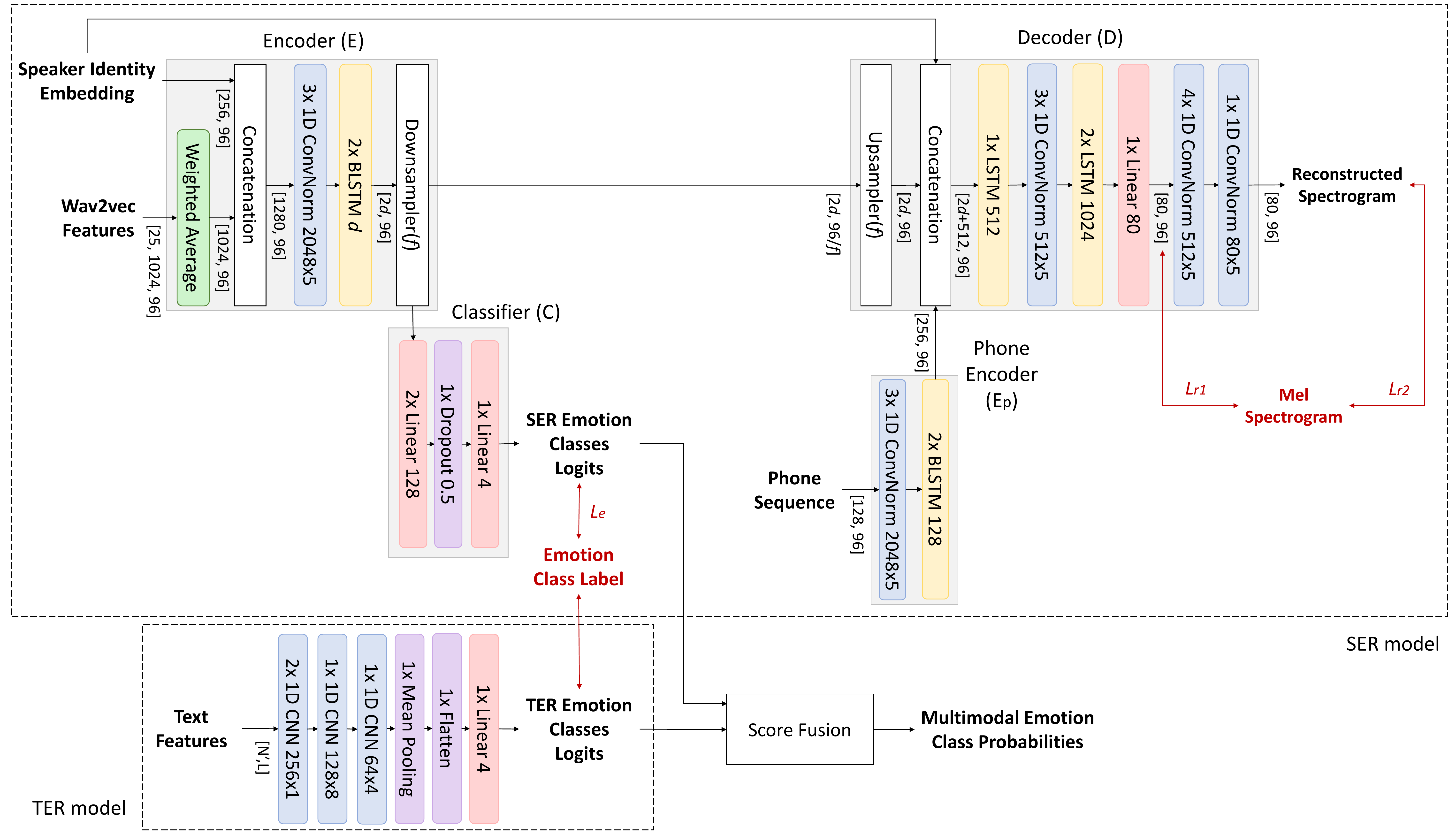}
  \caption{Proposed method depicting the SER and TER models and the fusion approach. 1D ConvNorm layers are defined as a 1D CNN layer followed by batch normalization, and BLSTM layers are bidirectional LSTMs. The number of layers is shown as each block name's prefix. The number of filters $F$ and kernels $K$ in each ConvNorm and CNN layers are shown as the block name's suffix $F\mathrm{x}K$. The number of neurons in LSTM, BLSTM, and Linear layers is shown as the blocks names' suffix.}
  \label{fig:model}
\end{figure*}

We propose a model to perform SER and a model to perform TER. The SER model takes as input wav2vec features, a mel spectrogram, speaker identity embeddings, and a phone sequence. All these features are extracted from the same speech segment, and the model outputs the probabilities for each emotion class, for the speech segment. The TER model takes as input text features extracted from an utterance's transcript, and outputs the probabilities for each emotion class, for the utterance. The SER and TER results are combined via score fusion to obtain the multimodal emotion class probabilities. Our proposed method, including the SER model, the TER model and the fusion approach, is depicted in Figure~\ref{fig:model}.

\subsection{Speech Emotion Recognition}
\label{sec:methodology-sub:ser}



We propose an encoder-decoder model that takes wav2vec 2.0 features as input, and reconstructs the corresponding mel-frequency spectrogram. Our model has four main components: an encoder, a decoder, a phone encoder, and a classifier. The model is trained over speech segments of $96$ frames, which is about $2$ seconds long. These segments are randomly cropped from the speech utterances during training.



Our SER model is similar to AutoVC~\cite{qian2019autovc}. However, our model differs in three aspects. First, wav2vec features are the acoustic input to our encoder. Second, we include an emotion classifier and an emotion loss to our method. Third, we define a phone encoder, whose output is inputted to the decoder.

\subsubsection{Wav2vec 2.0 Feature Extraction}
We extract the wav2vec features from a ``large'' wav2vec 2.0 model pre-trained on $60$k hours of unlabelled speech data from the LibriVox dataset\footnote{https://huggingface.co/facebook/wav2vec2-large-lv60}. We take the features from the feature encoder's output and from the output of all the $24$ Transformer layers in the context network. Thus, for each speech frame, there are $25$ $1024$-dimensional wav2vec features.


\subsubsection{Speaker Identity Feature Extraction}
We extract speaker identity features with Resemblyzer~\cite{wan2018generalized}\footnote{We use the code in: https://github.com/resemble-ai/Resemblyzer}, which is pre-trained on LibriSpeech~\cite{panayotov2015librispeech}, VoxCeleb1~\cite{nagrani17vox} and VoxCeleb2~\cite{chung18bvox}. For each utterance, a $256$-dimensional embedding is obtained to represent the speaker identity. For each speaker, we extract speaker identity features from $100$ randomly selected utterances, and we take their average as the final identity embedding to represent the speaker.

\subsubsection{Encoder}
A weighted average $\mathbf{h_{\mathrm{avg}}}$, with trainable weights $\alpha$, over the $25$ wav2vec 2.0 features $\mathbf{h_i}$, is computed as described in \cite{pepino2021emotion}:

\begin{equation}
    \mathbf{h_{\mathrm{avg}}} = \frac{\sum^{25}_{i=1}\alpha_i \mathbf{h_i}}{\sum^{25}_{i=1}\alpha_i}.
    \label{eq:weighted_avg}
\end{equation}

$\mathbf{h_{\mathrm{avg}}}$ is then concatenated with the $256$-dimensional speaker identity embedding frame by frame.


The BLSTM layers in the encoder have $d$ neurons, and their output have a size of $[2d, 96]$ since we concatenate the layers' output in both forward and backward directions. $d$ determines the size of the bottleneck, as it reduces the size of the features in the channel dimension. The downsampler operation~\cite{qian2019autovc} takes as input an array of size $2d$ for each speech frame, and returns the arrays taken at every $f$ frames. Thus, this operation reduces the temporal dimension of the feature array, by the downsampling factor $f$. The encoder outputs a feature array of size $[2d, 96/f]$, in which $d$ and $f$ control the bottleneck dimension. By controlling the size of the bottleneck, we would like to obtain a disentangled speech representation, that contains emotion information, but that does not contain speaker identity or phonetic information.

\subsubsection{Decoder}

At the decoder, the encoder's features are upsampled so that their size is the same as before the downsampling operation, by repeating each feature in the temporal dimension $f$ times. Since the encoder's features contain only emotion information, the decoder takes as input not only the encoder's output, but also speaker identity embeddings and phone sequence embeddings to be able to reconstruct the spectrogram.

The output from the decoder's linear layer is a feature array of size $80$ for each speech frame, which represents a mel-spectrogram of the speech segment. These features are compared with the ground-truth mel-spectrogram by means of a reconstruction loss $L_{r1}$, which is used to update all the model's parameters. We also compute the reconstruction loss $L_{r2}$ between the decoder's output and the same ground-truth mel-spectrogram. $L_{r1}$ and $L_{r2}$ are computed as

\begin{equation}
    L_r = \frac{1}{M}\sum_{k=1}^M(x_k - y_k)^2,
\end{equation}

\noindent in which $M$ is the training batch size, $x_k$ is the $k$-th feature element in the batch outputted by the model, and $y_k$ is the corresponding ground-truth for $x_k$.

\subsubsection{Phone Encoder}
The phone encoder takes as input a sequence of phone embeddings, and outputs a representation for the whole phone sequence. We follow two steps to obtain these phone embeddings. First, for each utterance, we extract the phone alignment information from the speech signal and its corresponding text transcript, by using the Gentle aligner\footnote{https://github.com/lowerquality/gentle}. Second, we obtain the phone sequence from the phone alignment information, by determining the longest phone for each frame. We define an id number to each phone, and we also assign ids to silence, not-identified phones, and to each special token in the dataset (e.g. ``[LAUGHTER]''), totalling $128$ distinct phone ids represented as one-hot embeddings.


\subsubsection{Classifier}
The classifier encourages the encoder's output to contain emotion information. We compute the cross-entropy loss $L_{e}$ between the emotion label $c$ and the softmax of the logits $z$ outputted by the classifier as

\begin{equation}
    L_e(z, c) = -\mathrm{log}\Bigg(\frac{\mathrm{exp}(z[c])}{\sum_j\mathrm{exp}(z_j)}\Bigg).
\label{eq:le}
\end{equation}

\subsubsection{Training and inference}

The objective function to be minimized during training is given as the sum between $L_{r1}$, $L_{r2}$ and $L_{e}$.


During inference, the softmax function is applied to the model's outputted emotion classes logits array to obtain the emotion class probabilities. The class with the highest probability is selected as the final classification result. The model takes as input features from a speech segment of $96$ frames. Thus, to obtain an utterance-level prediction, we first compute the emotion class probabilities every $96$ consecutive frames in an utterance (without overlap), zero-padding as necessary, and we take the average of these segment-level probabilities as the final utterance-level probability.


\subsection{Text Emotion Recognition}
\label{sec:methodology-sub:ter}


\cite{pepino2020fusion} shows that the TER task can benefit from processing embeddings of all text tokens before performing the emotion classification. Inspired by these results, we propose a CNN-based model to process all the token's embeddings in an utterance, extracted with Transformer-based models.


We extract a text representation of shape $[N, L]$ for each utterance with pre-trained Transformer-based models, in which $N$ is the number of tokens in the utterance excluding special tokens, and $L$ is the size of each token's feature. We zero-pad the text representation so that the input to the TER model have size $[N', L]$, in which $N'$ is the maximum number of tokens found in an utterance of the dataset. These text features are processed with the TER model illustrated in Figure~\ref{fig:model}, which is trained on the cross-entropy loss defined in Equation~(\ref{eq:le}). Similar to our SER model, during inference, the softmax function is applied to the TER model's logits array to obtain the emotion class probabilities. However, differently from the SER model, the output from the TER model represents the utterance-level emotion classification result.

\subsection{Multimodal Emotion Recognition}
\label{sec:methodology-sub:mer}


The speech-based utterance-level probabilities $p_s$ and the probabilities outputted from the text model $p_t$ for the same utterance are combined as

\begin{equation}
    p_f = w_1\cdot p_s + w_2\cdot p_t,
\end{equation}

\noindent in which $p_f$ is the fused probability, and $w_1$ and $w_2$ are fixed weights assigned to the speech and text modalities, respectively. The weights determine the degree of contribution of each data modality to the fused probability, and the emotion classification result for an utterance corresponds to the emotion class with the highest fused probability.

\section{Dataset}
\label{sec:dataset}


We utilize the Interactive Emotional Dyadic Motion Capture (IEMOCAP)~\cite{busso2008iemocap} dataset to evaluate our method. Given the amount of data and the phonetic and semantic diversity of its utterances, the IEMOCAP dataset is considered well-suited for speech-based and text-based emotion recognition.

There are $10$ actors in this dataset, whose interactions are organized in $5$ dyadic sessions, each with an unique pair of male and female actors. The dataset contains approximately $12$ hours of audiovisual data, which is segmented into speech turns (or utterances). Each utterance is labelled by three annotators.


Following previous works~\cite{krishna2020multimodal,yeh2020speech,feng2020end}, we consider only the utterances which are given the same label by at least two annotators, and we merge the utterances labelled as ``Happy'' and ``Excited'' into the ``Happy'' category. We further select only the utterances with the labels ``Angry'', ``Neutral'', ``Sad'' and ``Happy'', resulting in $5{,}531$ utterances, which is approximately $7$ hours of data. We utilize only the speech data, the transcripts, and the labels.

\section{Training Configuration}
\label{sec:train_config}

We perform a leave-one-session-out cross-validation in all our experiments. We report our results in terms of Weighted Accuracy (WA) and Unweighted Accuracy (UA). WA is equivalent to the average recall over all the emotion classes and UA is the fraction of samples correctly classified.

All models are implemented in PyTorch, and, in every training experiment, we use the Adam optimizer with learning rate $10^{-4}$, and with the default exponential decay rate of the moment estimates. The SER models are trained with a batch size of $2$ for $1$ million iterations. The TER models are trained with a batch size of $4$ for $412{,}800$ iterations.


\section{Speech Emotion Recognition}
\label{sec:ser}







\subsection{Emotion Recognition Experiments}
\label{sec:ser-sub:btneck}


We perform SER with two bottleneck configurations for the encoder, ``Small'' and ``Large'', which have respective bottleneck dimension $d$ equal to $8$ and $128$, and respective downsampling factor $f$ set as $48$ and $2$. The utterance-level SER results are presented in Table~\ref{tab:bttnk_res}.

\begin{table}[t]
  \centering
  \caption{UA (\%) results for the SER task on a model with a small bottleneck and a model with a large bottleneck.}\label{tab:bttnk_res}
  \begin{tabular}{|l|r|r|r|r|r|r|}
    \hline
    \multirow{2}{*}{Model} & \multicolumn{5}{c|}{Fold} & \multirow{2}{*}{Avg $\pm$ std} \\ \cline{2-6}
     &  1 &  2 &  3 &  4 &  5 &  \\ \hline
    Small & 67.9 & 71.7 & 67.3 & 72.2 & 71.6 & $\mathbf{70.1 \pm 2.3}$ \\ \hline
    Large & 59.6 & 73.6 & 66.0 & 71.3 & 70.0 & 68.1 $\pm$ 5.5 \\ \hline
  \end{tabular}
\end{table}

Table~\ref{tab:bttnk_res} indicates that the ``Small'' configuration performs better in the SER task when compared to the ``Large'' model. We compare the results obtained with the ``Small'' model with the current state-of-the-art in Table~\ref{tab:ser_comparison_cv}.

\begin{table}[t]
  \centering
  \caption{Comparison of our SER results with current works in terms of UA(\%) and WA (\%).}\label{tab:ser_comparison_cv}
  \begin{tabular}{|l|r|r|}
    \hline
    Model & UA & WA \\ \hline
    GRU+Context~\cite{rajamani2021novel} & 68.3 & 66.9 \\ \hhline{|=|=|=|}
    Self-Attn+LSTM~\cite{krishna2020multimodal} & 55.6 & -  \\ \hline
    BLSTM+Self-Attn~\cite{feng2020end} & 57.0 & 55.7 \\ \hline
    Transformer~\cite{zhang2021transformer} & - & 64.9 \\ \hline
    CNN+Feat-Attn~\cite{mao2020advancing} & 66.7 & - \\ \hline
    wav2vec+CNN~\cite{pepino2021emotion} & - & 67.9 \\ \hline
    Ours (Small) & \textbf{70.1} & \textbf{70.7} \\ \hline
  \end{tabular}
\end{table}







We further evaluate whether inputting the wav2vec embeddings is advantageous to SER. We train our model with the same parameters as the ``Small'' configuration, but with a mel-spectrogram as input instead of the wav2vec features. Our results achieved an UA of $50.4\%$ on the $5$-fold cross-validation, which is $19.7\%$ worse than of the model with wav2vec features as input, in terms of absolute accuracy. Therefore, we can conclude that the learned weighted average of the wav2vec embeddings is a better representation of speech for SER on the IEMOCAP dataset when compared to the traditional mel-frequency spectrograms.





\subsection{Disentanglement Experiments}
\label{sec:ser-sub:disent}


We train $4$-linear-layer (with, from the input to the output, $2048$, $1024$, $1024$, and $8$ neurons) classifiers on the obtained speech representations to solve the speaker identification task. Our goal is to see if the obtained emotion features contain speaker identity information. Ideally, we would like our features to be speaker-independent, and to hold a generic emotion representation that could be used across speakers.


We train the classifiers on a $5$-fold cross-validation, but we define the folds differently for this experiment. We randomly separate $80\%$ of each speaker's data for training, and the remaining $20\%$ for test. The folds have speaker dependent train and test sets, and each of them contains the data of only $4$ sessions. We train the classifiers on a cross-entropy loss. Table~\ref{tab:disent} summarizes the speaker identity recognition results.

\begin{table}[t]
  \centering
  \caption{UA (\%) results for the speaker identification task on features extracted with the ``Small'' and ``Large'' models.}\label{tab:disent}
  \begin{tabular}{|l|r|r|r|r|r|r|}
    \hline
    \multirow{2}{*}{Model} & \multicolumn{5}{c|}{Fold} & \multirow{2}{*}{Avg $\pm$ std} \\ \cline{2-6}
     & 1 & 2 & 3 & 4 & 5 &  \\ \hline
    Small & 18.4 & 20.6 & 19.4 & 15.9 & 13.6 & 17.6 $\pm$ 2.8 \\ \hline
    Large & 25.6 & 19.8 & 24.5 & 23.4 & 21.4 & 22.9 $\pm$ 2.3 \\ \hline
  \end{tabular}
\end{table}

Table~\ref{tab:disent} suggests that the features extracted with the ``Large'' model contain more information about speaker identity than the ones from the ``Small'' model. Overall, from the results in Tables~\ref{tab:disent} and \ref{tab:bttnk_res}, we can see that the features extracted with the ``Small'' model can achieve a better SER accuracy and a worse speaker identity accuracy when compared to the features extracted with the ``Large'' model. This result suggests that the bottleneck size can lead to a disentanglement of factors in speech, which makes the SER task easier.

\subsection{Discussion}

We believe that our SER model achieves better results when compared to previous methods due to three factors. First, we use high-level speech representations as the input to our model, which, apart from our work, is only done by \cite{pepino2021emotion} and \cite{zhang2021transformer}. Second, we are careful in analyzing the type of information encoded in the features obtained by our model, which makes the features have a certain level of disentanglement as shown in Section~\ref{sec:ser-sub:disent}. Third, our model can leverage both high-level and low-level features since it is trained to reconstruct spectrograms from wav2vec features.

\section{Text Emotion Recognition}
\label{sec:ter}




We train the TER model with input features extracted from different Transformer-based models\footnote{The trained models can be found at https://huggingface.co/models}. We use the ``large'' version of all these models, which output a $1024$-dimensional feature for each token. The TER results are shown in Table~\ref{tab:txt_acc} for different feature extractors, and we compare our best results with the current state-of-the-art in Table~\ref{tab:ter_comparison_cv}.








\begin{table}[t]
  \centering
  \caption{5-fold cross-validation UA (\%) results for TER on input features extracted from different models. ($c$ = ``cased'', $u$ = ``uncased'', $uwm$ = ``uncased with whole word masking'')}\label{tab:txt_acc}
  \begin{tabular}{|l|r|}
    \hline
    {Model} & {Avg $\pm$ std} \\ \hline
    ALBERT & 62.3 $\pm$ 2.3\\ \hline
    BERT$_{c}$ & 65.5 $\pm$ 3.3\\ \hline
    BERT$_{u}$ & $\mathbf{66.1 \pm 2.1}$ \\ \hline
    BERT$_{uwm}$ & 65.8 $\pm$ 2.6\\ \hline
    ELECTRA & 56.6 $\pm$ 3.8\\ \hline
    RoBERTa & 64.1 $\pm$ 3.5\\ \hline
    XLNet$_{c}$ & 58.1 $\pm$ 3.6 \\ \hline
  \end{tabular}
\end{table}



\begin{table}[t]
  \centering
  \caption{Comparison of our TER results with current works in terms of UA(\%) and WA (\%).}\label{tab:ter_comparison_cv}
  \begin{tabular}{|l|r|r|}
    \hline
    Model & UA & WA \\ \hline
    BERT+Attn+Context~\cite{wu2021emotion} & 71.9 & 71.2 \\ \hhline{|=|=|=|}
    BERT+Attn~\cite{wu2021emotion} & 64.8 & 62.9 \\ \hline
    BLSTM+Self-Attn~\cite{feng2020end} & 63.6 & 63.7 \\ \hline
    Self-Attn+LSTM~\cite{krishna2020multimodal} & 65.9 & -  \\ \hline
    Ours (BERT uncased) & \textbf{66.1} & \textbf{67.0} \\ \hline
  \end{tabular}
\end{table}



From Table~\ref{tab:ter_comparison_cv}, we can see that our method achieves better results than previous works, except for \cite{wu2021emotion}, which uses context information (i.e., features from succeeding and preceding utterances). Our model differs from previous works in that we do not use recurrent neural networks or self-attention, and we benefit from the text representations learned by Transformer-based models trained on large text corpora. We believe our model achieved good results due to BERT's deep features, and to the ability of our model's 1D CNN layers to extract temporal information from the sequence of token's features.

\section{Multimodal Emotion Recognition}
\label{sec:mer}


We combine the results from our best speech and text models by experimenting with different weight values $w_1$ and $w_2$. Our best multimodal results are acquired when $w_1=0.6$ and $w_2=1$, and they are reported in Table~\ref{tab:mer_comparison_cv}.

\begin{table}[t]
  \centering
  \caption{Comparison of our multimodal results with current works in terms of UA(\%) and WA (\%).}\label{tab:mer_comparison_cv}
  \begin{tabular}{|l|r|r|}
    \hline
    Model & UA & WA \\ \hline
    BERT+Attn+Context~\cite{wu2021emotion} & 76.1 & 77.4 \\ \hhline{|=|=|=|}
    LAS-ASR~\cite{yeh2020speech} & 66.0 & 64.0 \\ \hline
    ASR-SER~\cite{feng2020end} & 69.7 & 68.6 \\ \hline
    CMA+Raw waveform~\cite{krishna2020multimodal} & 72.8 & -  \\ \hline
    Ours ($w_1=0.6$, $w_2=1$) & \textbf{73.0} & \textbf{73.5} \\ \hline
  \end{tabular}
\end{table}

This result shows that, when combining the speech and the text results, it is better to give less importance to the speech model's result, even though its accuracy is higher than the text model's. We believe this may be related to the confidence in which the TER and the SER models obtain their scores, but further investigation is required.

By comparing our results in Tables~\ref{tab:bttnk_res}, \ref{tab:txt_acc}, and \ref{tab:mer_comparison_cv}, we can conclude that the solution to the emotion recognition task benefits from combining different types of data, since our multimodal result is better than our speech-only and text-only results. Our multimodal approach gives better results than current works except for \cite{wu2021emotion}, which uses the context information. We attribute the reason of our good results to the fact that our unimodal models outperform other unimodal models, and not to our choice of fusion method. We believe we could achieve better results with a more sophisticated fusion approach or by jointly training the speech and text modalities.

\section{Conclusion}
\label{sec:conclusion}


We proposed a cross-representation encoder-decoder model inspired in disentanglement representation learning to perform SER. Our model leverages both high-level wav2vec features and low-level mel-frequency spectrograms, and it achieves an accuracy of $70.1\%$ on the IEMOCAP dataset. We also used a CNN-based model that processes token's embeddings extracted with pre-trained Transformer-based models to perform TER, achieving an accuracy of $66.1\%$ on the same dataset. We further combined the speech-based and the text-based results via score fusion, achieving an accuracy of $73.0\%$. Our speech-only, text-only and multimodal results surpassed current works', showing that emotion recognition can benefit from disentanglement representation learning, high-level data representations, and multimodalities.




%


\bibliographystyle{IEEEbib}
\bibliography{bibliography}

\end{document}